\documentclass[aps,pra,reprint,superscriptaddress, nofootinbib]{revtex4-2}
\usepackage[LGR,T1]{fontenc}
\usepackage[utf8]{inputenc}
\usepackage{lipsum}
\usepackage{comment}
\usepackage{graphicx}
\usepackage{babel}
\usepackage{amsmath,amssymb, amsthm}
\usepackage{float} 
\usepackage{caption, subcaption}
\usepackage{enumitem}
\usepackage{esdiff}
\usepackage{mathrsfs}
\usepackage{mathtools}
\usepackage{enumitem}
\usepackage{braket}
\usepackage{bbm}   %For mathbbm{1}
\usepackage{hyperref}
\usepackage{tikz}
\usepackage{ytableau} 
\usepackage{csquotes}%enquote
\usepackage{verbatim}

\usetikzlibrary{positioning}
\usetikzlibrary{calc}

\def\*#1{\boldsymbol{#1}} %abbreviazione scritta grassetto
\def\.#1{\text{#1}} %abbreviazione testo in math mode
\def\>#1{\mathlarger{#1}}
\def\->#1{\overrightarrow{#1}}
\def\##1{\hat{#1}}
\def\bra#1{{\langle{#1}|}}
\def\ket#1{{|{#1}\rangle}}

\begin{document}

\title{Measurement-induced dynamics and emergent symmetries of particles moving in a one-dimensional lattice}

\author{Salvatore Di Lorenzo}
\email{sdl@phys.au.dk}
\address{Department of Physics and Astronomy, Aarhus University, Universitetsparken 524, 8000 Aarhus, Denmark}

\author{Klaus Mølmer}
\email{klaus.molmer@nbi.ku.dk}
\address{Niels Bohr Institute, University of Copenhagen, Jagtvej 155, 2200 Copenhagen, Denmark}

\date{\today}

\begin{abstract}
    Continuous measurements simultaneously reveal and modify the dynamics of a quantum system. In this work we consider the tunneling motion of particles between the sites of a one-dimensional lattice. We investigate how weak continuous probing of the occupation of a single lattice  site induces entanglement and selects definite values for system properties such as wave function parity and permutation symmetry. Our simulations show how continuous measurement can steer systems of two and three particles, with no prior permutation symmetry, into stable bosonic, fermionic and parastatistical symmetry sectors.
\end{abstract}

\maketitle

\section{Introduction}
In the standard quantum mechanical description of measurements, the outcome is random and restricted to the eigenvalues of the measured observable and the state is projected onto the corresponding eigenstate. 

Not all measurements are characterized by a direct projection on an eigenstate basis for a system observable. In fact, many real measurements take place by interacting a quantum system with a quantum meter system, and measurements on the meter system cause the system of interest to evolve by the more general formalism of positive operator valued measures (POVM) \cite{qip_bergou}.
Examples include shining off-resonant laser light on atoms in optical lattices and detecting the light by photon counting \cite{bakr2009,sherson2010} or by homodyne detection of the probe light, as in Faraday probing \cite{yamamoto2017,hammerer2010}. In the case of continuous probing of a system with a beam of light, homodyne detection yields a stochastic signal governed by the mean value of the probed observable $X$ and a Gaussian noise $\.dW$ \cite{Jacobs2014}. These dynamics are described by the stochastic Schrödinger equation (SSE):
\begin{equation}
\begin{aligned}
    &\.d\ket{\psi(t)} = \Big\{ - \frac{i}{\hbar}H\.dt -k(X-\braket{X})^2 \.dt +\\
    &+ \sqrt{2k}(X-\braket{X})\.dW \Big\}\ket{\psi(t)}
    \label{eq:SSE}
\end{aligned}
\end{equation}
for pure state dynamics, and the stochastic master equation (SME):
\begin{equation}
\begin{aligned}
    &\.d\rho = - \frac{i}{\hbar}[H,\rho]\.dt  - k[X,[X,\rho]]\.dt + \\
    &\sqrt{2k}(X\rho + \rho X - 2\braket{X}\rho)\.dW
\end{aligned}
\label{eq:SME}
\end{equation}
for mixed state dynamics. The parameter $k$ denotes the \emph{measurement strength} and depends on the intensity of the probe field and the coupling strengths between the field and system of interest. 

Previous works have explored the richness of continuous measurement dynamics in several systems \cite{anton-molmer, cao2019, blattmann2016} and in particular, how the weak probing of an observable $X$ can indirectly provide information on another unprobed observable $A$ (such as photon number in a cavity, or parity of the quantum state of a particle moving in a lattice) and steer a system into subspaces where conserved quantities attain definite values. 

This can be illustrated by extending the Ehrenfest theorem:
\begin{equation}
    \diff{\braket{A}}{t} = -\frac{i}{\hbar}\braket{[A,H]} + \left\langle{\diffp{A}{t}}\right\rangle
    \label{eq:ehrenfest_standard}
\end{equation}
to systems subject to probing, using the SME:
\begin{equation}
\begin{aligned}
    \.d\braket{A} =& \ \.{d Tr}(\rho A) =  \.{Tr}(\.d\rho\, A) +  \.{Tr}(\rho \,\.d A) \\
    =&-\frac{i}{\hbar}\braket{[A,H]}\.dt - k\braket{[X,[X,A]]}\.dt\\
    &+ \sqrt{2k}\braket{AX+XA-2\braket{X}A}\.dW + \braket{\partial_t A}
\end{aligned}
\label{eq:measured_ehrenfest_theorem}
\end{equation}
where the explicit time dependence $\braket{\partial_t A}$ vanishes for time-independent observables. We note that, while the signal remains noisy, the term proportional to $\.dW$ vanishes if the system is in an eigenstate or eigenspace of the operator $A$, and if $[A,H]=0$ and $[X,[X,A]]=0$, the dynamics preserves such eigenspaces \cite{anton-molmer}. 

In this work, we explore this convergence process with one, two and three particles moving in a one-dimensional lattice. We simulate the system dynamics through the stochastic Schrödinger equation, using the solver \verb|ssesolve| from the Python package QuTiP\cite{qutip5}.

Formally, if $A$ has a finite spectrum, the time evolution is a bounded martingale, and the convergence of $\braket{A}$ to a steady state for $t\longrightarrow\infty$ is guaranteed by Doob's Theorem \cite{doob1953}. Therefore, the continuous measurement of $X$ causes the projection of a quantum system into an eigenspace of an observable $A$ that commutes with the Hamiltonian $H$ and the probed observable $X$.

Due to the continuous probing of the occupation of a single lattice site, we infer values of the parity and the exchange symmetry of the particles. Elementary and composite particles in nature are bosons or fermions with given prior exchange symmetries, but we can make a model where different particles are practically indistinguishable by our measurements, and we show that weak probing of the total occupation of a single spatial location can ultimately project such systems into the symmetry classes governing bosons or fermions, \emph{or} alternative para-statistical symmetries, which are not observed for any elementary particles in nature.

\section{One particle dynamics in a lattice}
The dynamics of a single particle in a discrete lattice subject to continuous weak probing has been investigated in Ref. \cite{blattmann2016}.
Following the Hubbard model \cite{jakschzoller2005}, the state can be expanded in the basis $\{\ket{n}\}$, describing a particle populating L sites, labeled as $n=1,...,L$. The kinetic energy operator is associated with the tunneling between neighboring sites and is given by the tight-binding Hamiltonian \cite{valiente2010}:
\begin{equation}
    H = -J \sum_{n=1}^{L-1} (\ket{n}\bra{n+1} + \ket{n+1}\bra{n})
    \label{eq:tight_binding_hamiltonian_1part}
\end{equation}
with the energy eigenstates
\begin{equation}
    \ket{E_k} = \sqrt{\frac{2}{L+1}}
    \sum_{n=1}^{L} \sin\left( \frac{\pi k n}{L+1} \right) \ket{n}
    \label{eq:energy_eigenstates}
\end{equation}
and eigenvalues $E_k = -2J\cos\left(\frac{\pi k}{L+1}\right)$, for $k=1,...,L$.

\begin{figure}[H]
    \centering
    \hspace{0.2cm}\includegraphics[width=0.48\linewidth]{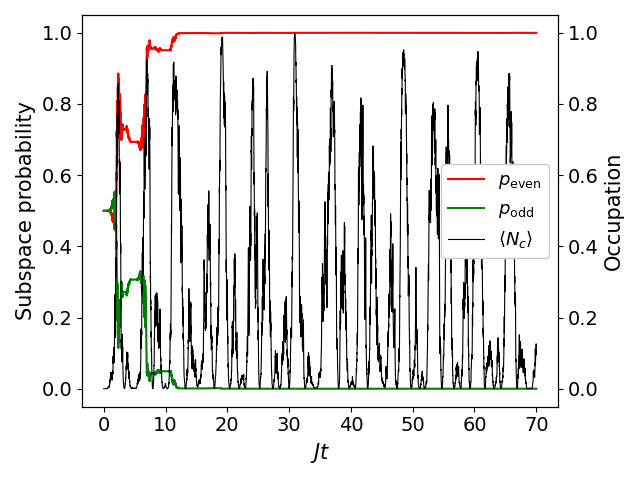}
    \includegraphics[width=0.48\linewidth]{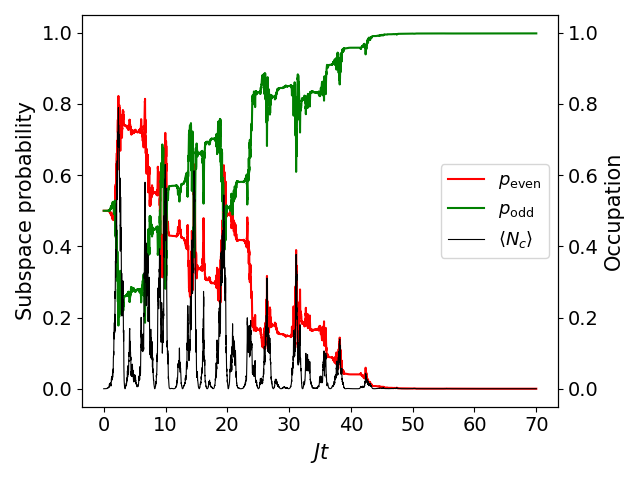}\\
    \hspace{-0.15cm}\includegraphics[width=0.43\linewidth]{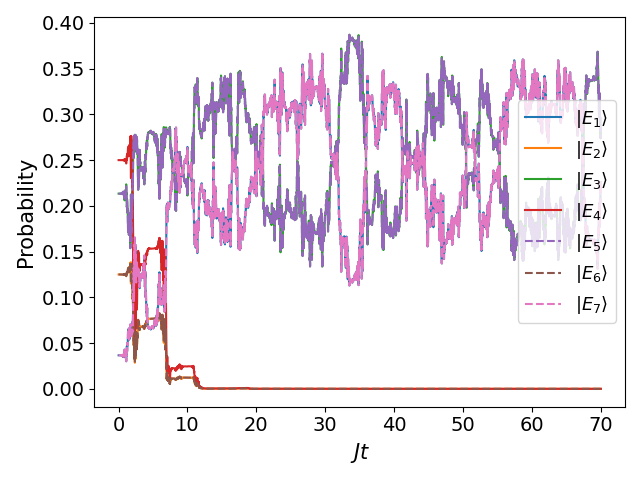}\hspace{0.5cm}
    \includegraphics[width=0.43\linewidth]{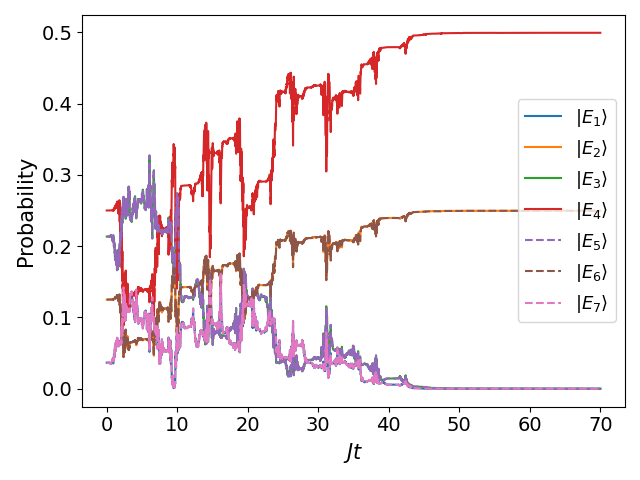}
    \caption{The upper panels show the mean population of the probed, middle site, and the ensuing selection of the even and odd subspace, from the initial state $\ket{\psi(0)}=\ket{1}$. The lower panels show the corresponding evolution of the energy eigenstate populations. The even subspace shows internal fluctuations between different eigenstates, whereas in the odd subspace the ratios between the populations remain constant.}
    \label{fig:1part}
\end{figure}

We now consider a lattice with $L=7$ and a particle initially localized at $n=1$. Since the number of sites is odd, we can define the central site $c=4$, and probe the occupation number $\#N_c=\ket{c}\bra{c}$.
\begin{figure}[H]
    \centering
    \begin{center}
    \begin{tikzpicture}
    \def\site{0.7}
    \def\d{0.5}
    \def\dd{0.6}
    \colorlet{meas_color}{red!70!black}
    
    %First two sites
    \draw[-, thick] (0,0) -- (\site,0) ;
    \node[above] at (0.5*\site,0) {$\ket{1}$};
    \draw[-, thick](\site+\d,0) -- (2*\site+\d,0);
    \node[above] at (1.5*\site+\d,0) {$\ket{2}$};

    \draw[fill=black] (0.5*\site,0) circle (0.1cm);
    
    %Center
    \node at (2*\site+\d+\dd,0) {$\cdots$};
    \draw[-, thick, meas_color] (2*\site+\d+2*\dd,0) -- (3*\site+\d+2*\dd,0);
    \node[above] at (2.5*\site+\d+2*\dd,0) {$\ket{c}$};
    \node at (3*\site+\d+3*\dd,0) {$\cdots$};

    %End
    \draw[-, thick] (3*\site+\d+4*\dd,0) -- (4*\site+\d+4*\dd + 0.3,0);
    \node[above] at (3.5*\site+\d+4*\dd+0.15,0) {$\ket{L-1}$};

    \draw[-, thick] (4*\site+\d+5*\dd +0.3,0) -- (5*\site+\d+5*\dd +0.3,0);
    \node[above] at (4.5*\site+\d+5*\dd+ 0.3,0) {$\ket{L}$};

    %Arrow
    \draw[->, ultra thick, meas_color, >=stealth] (2.5*\site+\d+2*\dd,-1) -- ++(0,0.8);
    \node[right] at (2.5*\site+\d+2*\dd,-0.7) {\small$\hat{N}_p = \ket{p}\bra{p}$};

    \draw[<->,  thick, out=-90, in=-90, looseness=0.75]
    (0.5*\site,-0.1) to (1.5*\site+\d,-0.1);
    \node[below] at (\site+0.5*\d, -0.4) {$J$};
\end{tikzpicture}
\end{center}
    \caption{Probing scheme for the one-particle chain.}
    \label{fig:placeholder}
\end{figure}
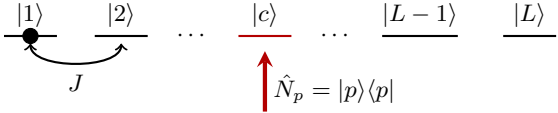
The initial state $\ket{\psi(0)}=\ket{1}$, located on the left, can be written as an equal superposition of an even and an odd state. The occupation of the central site commutes with the parity operator and the probing gradually transforms the state into one with well-defined even or odd parity, with equal probability. This is shown in the simulations depicted in Fig.\ref{fig:1part}, where the black curves in the upper panels show two examples of the mean occupation of the middle site $c=4$, simulated  by the SSE \eqref{eq:SSE} with a medium probing strength ($k=0.6 J$)\footnote{Different probing strengths only affect the convergence speed, without any qualitative impact of interest for our observations. In the simulations later shown, we choose $k$ between $0.6J$ and $0.8J$.}.

We note that the system may still evolve dynamically due to its occupation of superposition states in the odd and even  subspaces. In particular, the even subspace is still subject to noise due to the measurements on the central site, and the populations of the different energy eigenstates keep evolving, while the odd eigenstates populations are not affected by the measurement, and their relative probabilities are unchanged throughout the process (while their total probability changes as the system gradually attains the odd or even symmetry).

\section{Two particles}
\newcommand{\twoparticleschemeA}{\begin{center}
    \begin{tikzpicture}
    \def\site{0.7}
    \def\d{0.5}
    \def\dd{0.6}
    \def\h{0.3}
    \colorlet{meas_color}{red!70!black}
    
    %First two sites
    \draw[-, thick] (0,0) -- (\site,0) ;
    \node[above] at (0.5*\site,\h) {$\ket{1}$};
    \draw[-, thick](\site+\d,0) -- (2*\site+\d,0);
    \node[above] at (1.5*\site+\d,\h) {$\ket{2}$};

    \draw[fill=black] (0.5*\site,0) circle (0.1cm);
    \draw[->] (0.3*\site,-0.35) -- (0.7*\site, 0.35);
    
    %Center
    \node at (2*\site+\d+\dd,0) {$\cdots$};
    \draw[-, thick, meas_color] (2*\site+\d+2*\dd,0) -- (3*\site+\d+2*\dd,0);
    \node[above] at (2.5*\site+\d+2*\dd,\h) {$\ket{p}$};
    \node at (3*\site+\d+3*\dd,0) {$\cdots$};

    %End
    \draw[-, thick] (3*\site+\d+4*\dd,0) -- (4*\site+\d+4*\dd,0);
    \node[above] at (3.5*\site+\d+4*\dd,\h) {$\ket{L}$};
    \draw[fill=black] (3.5*\site+\d+4*\dd,0) circle (0.1cm);
    \draw[->] (3.7*\site+\d+4*\dd,-0.35) -- (3.3*\site+\d+4*\dd, 0.35);

     \draw[->, ultra thick, meas_color, >=stealth] (2.5*\site+\d+2*\dd,-1) -- ++(0,0.8);
    \node[right] at (2.5*\site+\d+2*\dd,-0.7) {\small$\hat{N}_p$};
\end{tikzpicture}
\end{center}}

\newcommand{\twoparticleschemeB}{\begin{center}
    \begin{tikzpicture}
    \def\site{0.7}
    \def\d{0.5}
    \def\dd{0.6}
    \def\sep{-1}
    \colorlet{meas_color}{red!70!black}
    
    %First two sites
    \draw[-, thick] (0,0) -- (\site,0) ;
    \node[above] at (0.5*\site,0) {$\ket{1}$};
    \draw[-, thick](\site+\d,0) -- (2*\site+\d,0);
    \node[above] at (1.5*\site+\d,0) {$\ket{2}$};

    \draw[fill=black] (0.5*\site,0) circle (0.1cm);
    
    %Center
    \node at (2*\site+\d+\dd,0) {$\cdots$};
    \draw[-, thick, meas_color] (2*\site+\d+2*\dd,0) -- (3*\site+\d+2*\dd,0);
    \node[above] at (2.5*\site+\d+2*\dd,0) {$\ket{p}$};
    \node at (3*\site+\d+3*\dd,0) {$\cdots$};

    %End
    \draw[-, thick] (3*\site+\d+4*\dd,0) -- (4*\site+\d+4*\dd,0);
    \node[above] at (3.5*\site+\d+4*\dd,0) {$\ket{L}$};

    %---- SECOND LATTICE ----
    
    %First two sites
    \draw[-, thick] (0,\sep) -- (\site,\sep) ;
    \draw[-, thick](\site+\d,\sep) -- (2*\site+\d,\sep);
    
    %Center
    \node at (2*\site+\d+\dd,\sep) {$\cdots$};
    \draw[-, thick, meas_color] (2*\site+\d+2*\dd,\sep) -- (3*\site+\d+2*\dd,\sep);
    \node at (3*\site+\d+3*\dd,\sep) {$\cdots$};

    %End
    \draw[-, thick] (3*\site+\d+4*\dd,\sep) -- (4*\site+\d+4*\dd,\sep);
    \draw[fill=black] (3.5*\site+\d+4*\dd,\sep) circle (0.1cm);

    %Arrow
    \draw[->, ultra thick, meas_color, >=stealth] (2.5*\site+\d+2*\dd,-0.5+\sep) -- ++(0,1.5);
    \node[right] at (2.5*\site+\d+2*\dd,-0.7+\sep) {\small$\hat{N}_p$};
\end{tikzpicture}
\end{center}}

The one-particle case admits different possible extensions to two and more particles. One may for example consider distinguishable atoms that evolve on the same lattice with different tunneling parameters and which may be separately probed; or the opposite situation of  indistinguishable atoms, for which the bosonic or fermionic symmetry is already enforced.

We consider, instead, a case in which the atoms are identical to the observer, while being in principle distinguishable. 
The atoms therefore have the same tunneling constant $J$, and the measurement on a site $p$ is only able to count the number of atoms:
\begin{equation}
    \#N_p = \ket{p}\bra{p}\otimes\mathbbm{1} + \mathbbm{1}\otimes \ket{p}\bra{p}
\end{equation}
The degrees of freedom that distinguish the atoms must, however, be inaccessible to the probe.

We suggest two possible implementations of this situation with practically indistinguishable  but formally different particles with atoms in optical lattices:
\begin{enumerate}
    \item We can use alkaline earth atoms with $J=0$, prepared in different magnetic  substates  of the nuclear spin (\cite{Gorshkov2010}), in which the nuclear spin orientation is decoupled from the Hamiltonian and a probe laser can only detect the presence of atoms but not their spin state (fig. \ref{fig:same_lattice}).
    \begin{figure}[H]
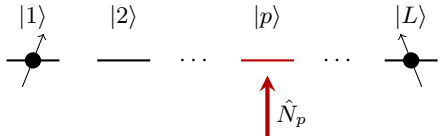

        \centering
        \twoparticleschemeA
        \caption{Same-lattice model for indistinguishable particles in the state $\ket{1}\ket{L}$. The arrows indicate that the two atoms populate different internal, nuclear spin states that are, however, not resolved by the measurement.}
        \label{fig:same_lattice}
    \end{figure}
    \item Alternatively, we can use a parallel arrangement of identical 1D lattices, each containing only one atom, and subject to probing of the total population of any site $\ket{p}$ in all lattices (fig. \ref{fig:parallel_lattices}).
    \begin{figure}[H]
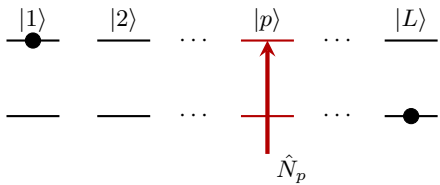

        \centering
        \twoparticleschemeB
        \caption{Parallel lattices occupied by single atoms (shown here in the product state $\ket{1}\ket{L}$). The probe beam crosses both lattices and does not distinguish between the atoms.}
        \label{fig:parallel_lattices}
    \end{figure}
\end{enumerate}

%Both of the above methods can be extended to more than two particles. With the method 1 it is possible to achieve up to 10 distinguishable atoms (for isotopes with spin $I=\frac{9}{2}$), while method 2 allows for even more, with the only limit being the creation and control of the optical lattice.

In the first scenario, atoms may interact when they occupy the same lattice site, which makes it harder to isolate the effects of probing from those of interactions. This is not an issue in the second scenario, where the atoms occupy displaced  lattices, and here the probe laser beam should be carefully aligned to detect the number of atoms intersecting the probe beam without resolving which lattices they occupy. 

We have simulated the dynamics of interacting atoms and found that, whereas parity does not in general converge to definite values in the presence of on-site interactions, the permutation symmetry still does\footnote{This can be seen as a consequence of the fact that the two-particle interaction operator $\#V = U \sum_n \ket{n}\bra{n}\otimes\ket{n}\bra{n}$ commutes with the permutation symmetry projectors introduced later. Therefore, because of the extended Ehrenfest theorem \eqref{eq:measured_ehrenfest_theorem}, we expect the symmetry projectors to converge to  definite values regardless of interacting potentials.}.

Focusing this work on the emergent exchange symmetry, we proceed here by considering non-interacting particles, and we will refer to the displaced lattice scheme for convenient discussion of the results.

\subsection{Entanglement and emergent symmetry}
%We note that identical atoms in the same internal state are by definition indistinguishable and their spatial states are thus symmetric or antisymmetric depending on the bosonic or fermionic nature of the atom. [Maybe redundant?]
We want to show in the following that permutation symmetry can appear as an emergent property, and to this end we simulate a situation of non-interacting particles with identical properties. We refrain from enforcing permutation symmetry in the initial state but assume instead an initial product of single-particle states,
\begin{equation} \label{pstate}
    \ket{\psi} = \ket{\psi_a}\otimes\ket{\psi_b}
\end{equation}
The Hamiltonian acts identically (while separately) on the two particles
\begin{equation}
\begin{aligned}
    &\#H = -J \sum_{n=1}^{L-1}(\ket{n}\bra{n+1}\otimes \mathbbm{1} +\mathbbm{1}\otimes \ket{n}\bra{n+1}) + \.{h.c.}
\end{aligned}
\end{equation}

The spatial motion of the atoms becomes entangled by the probing laser, similarly to other measurement schemes that enable interaction-free entanglement between atoms \cite{cabrillo1999}.
Given atoms $1$ and $2$ on the two 1D sub-lattices, populating a pure joint state $\Psi_{12}$ and the corresponding joint density matrix $\rho_{12}$, we quantify the entanglement between the two systems by the von Neumann entanglement entropy
\begin{equation}
    \mathcal{E} = -\.{Tr}_1 [\rho_{1}\log_2 \rho_{1}]
    \label{eq:von_neumann_entropy}
\end{equation}
where $\rho_1 = \.{Tr}_2(\rho_{12})$. If the atoms are unentangled, then $\mathcal{E}=0$; otherwise, $\mathcal{E}>0$.

Together with the entanglement comes the possibility for the particles to populate states of the form
\begin{equation} \label{pstate}
    \ket{\psi} = {\cal N}(\ket{\psi_a}\otimes\ket{\psi_b} \pm \ket{\psi_b}\otimes\ket{\psi_a}) 
\end{equation}
where the first and second atom is described by the left and right parts of the tensor product and ${\cal N}$ is a normalization constant.  In these states, the system has formally acquired the respective bosonic and fermionic symmetries.
The symmetries can be expressed in terms of the projection onto the fermionic and bosonic subspaces,
\begin{subequations}
\begin{equation}
    \hat{\mathcal{P}}_\.{bos} = \frac{1}{2}(1+\hat{P}_{12})
    \label{eq:P_bos_2part}
\end{equation}
\begin{equation}
    \hat{\mathcal{P}}_\.{ferm} = \frac{1}{2}(1-\#P_{12})
    \label{eq:P_ferm_2part}
\end{equation}
\end{subequations}
where we have defined the particle exchange operator $\hat{P}_{12}$,
\begin{equation}
    \hat{P}_{12} \ket{j}\otimes\ket{k} = \ket{k}\otimes\ket{j}.
\end{equation}

The expectation values of $\hat{\mathcal{P}}_\.{bos}$ ($\hat{\mathcal{P}}_\.{ferm}$) quantify the degree of symmetry (antisymmetry) of the state. For an initial product state of two orthogonal single particle states, $\braket{\mathcal{\#P}_\.{bos}} = \braket{\mathcal{\#P}_\.{ferm}} = \frac{1}{2}$. The expectation values of the projection operators fluctuate due to probing, but since $[\#H,\#P_{12}]=0$ and $ [\#N_p,\#P_{12}]=0$, we expect them to eventually reach a steady state. This prediction, which follows our reasoning on eq. \eqref{eq:measured_ehrenfest_theorem}, is confirmed by our simulations.

\begin{figure}[!t]
    \centering
    \begin{subfigure}{0.45\linewidth}
        \includegraphics[width=\linewidth]{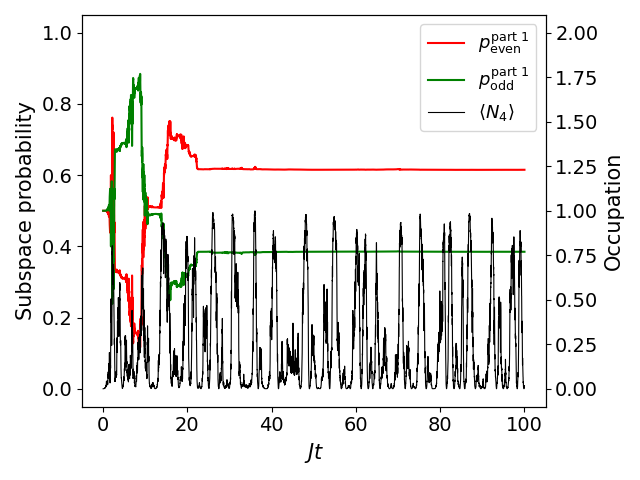}
    \end{subfigure}
    \begin{subfigure}{0.45\linewidth}
        \includegraphics[width=\linewidth]{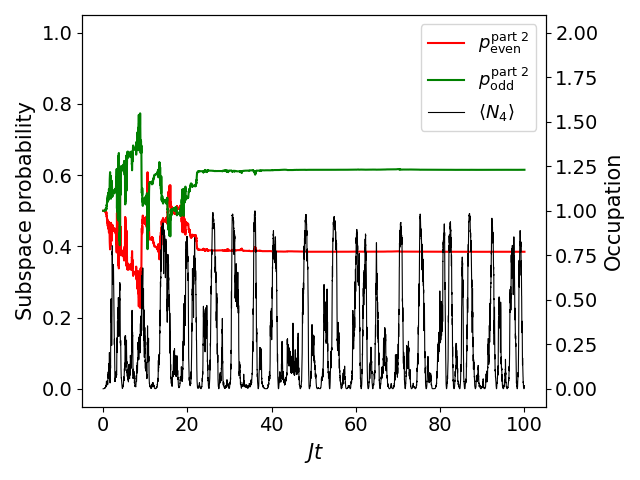}
    \end{subfigure}
    \begin{subfigure}{0.45\linewidth}
        \includegraphics[width=\linewidth]{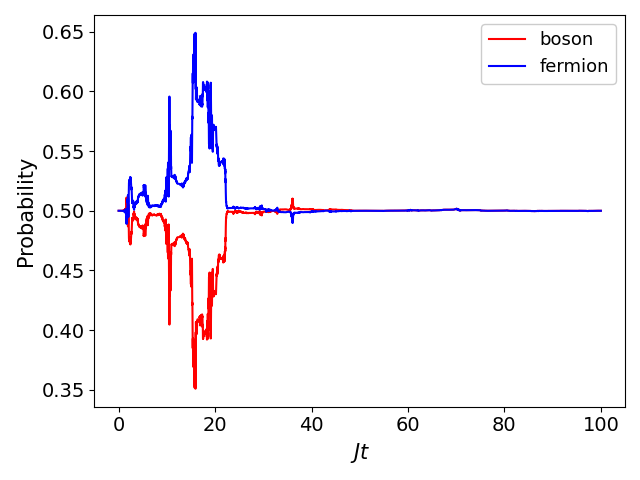}
    \end{subfigure}
    \begin{subfigure}{0.45\linewidth}
        \includegraphics[width=\linewidth]{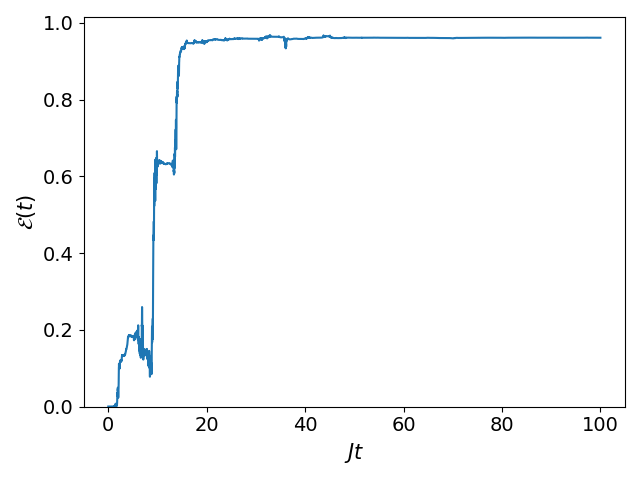}
    \end{subfigure}
    \caption{Selection of the EO parity subspace, from $\ket{\psi(0)}=\ket{1}\ket{2}$, by probing the central site $c=4$. The particles become entangled as a result of the incapability of measurement to distinguish which particle is in the even state or odd state. Nothing can be inferred about exchange symmetry, which remains half-bosonic half-fermionic.} 
    \label{fig:evenodd}
\end{figure}

By simulating a system with odd number of sites ($L=7$), we see again even and odd parity states. However, our measurement scheme cannot distinguish which atom is in the odd or even subspace, and dynamics of the middle site population can only reveal if they are both odd (OO), both even (EE), or one even and one odd (EO). The latter case typically presents itself with states of the form $\alpha\ket{\phi_\.{even}}\ket{\psi_\.{odd}} +\beta \ket{\psi_\.{odd}}\ket{\phi_\.{even}}$, such as in figure \ref{fig:evenodd}.
For the OO and EO cases, there is either no signal that depends on the atoms (OO) or the atoms populate orthogonal states (EO), and their symmetry under permutation cannot be inferred from the signal. Figure \ref{fig:evenodd} shows an example of EO state, where the exchange symmetry eventually settles to half-bosonic, half-fermionic. 

\begin{figure}[!t]
    \centering
    \begin{subfigure}{0.49\linewidth}
        \includegraphics[width=\linewidth]{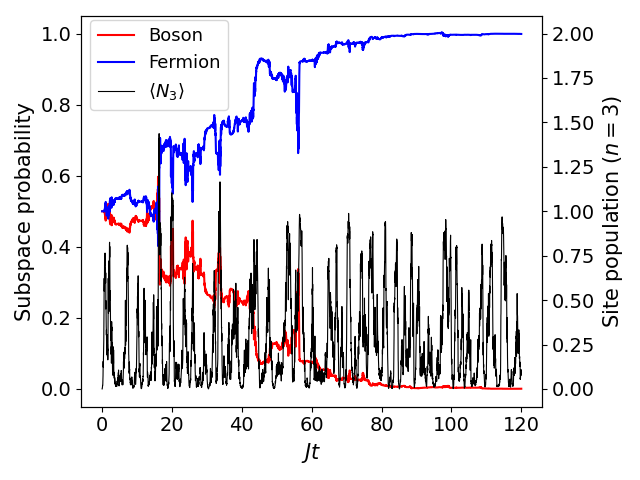}
    \end{subfigure}
    \begin{subfigure}{0.49\linewidth}
        \includegraphics[width=\linewidth]{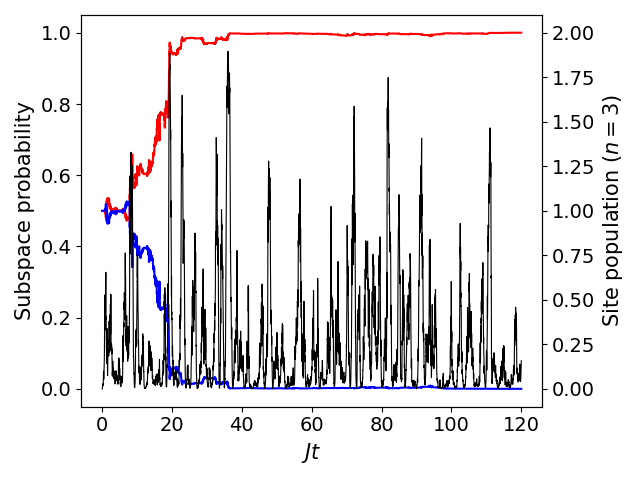}
    \end{subfigure}
    \begin{subfigure}{0.49\linewidth}
        \includegraphics[width=\linewidth]{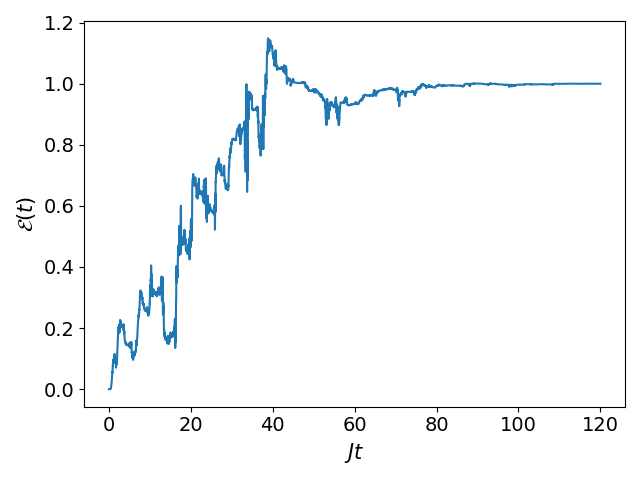}
    \end{subfigure}
    \begin{subfigure}{0.49\linewidth}
        \includegraphics[width=\linewidth]{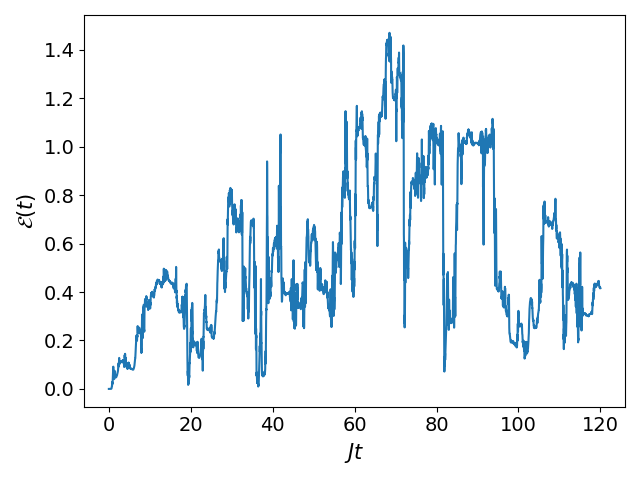}
    \end{subfigure}
    \caption{Selection of fermionic (left) and bosonic (right) states, from the initial product of position eigenstates $\ket{\psi}=\ket{n_1=1}\ket{n_2=2}$ by probing of the non-symmetric site $p=3$ on a lattice with $L=7$ sites.}
    \label{fig:noncentral_01_ferm}
\end{figure}

However, when both particles \emph{can} simultaneously occupy the probing site, such as in the EE case  or when a non-central site is probed, then any initial product state is unambiguously steered towards a bosonic or fermionic final state.
This is demonstrated in figure \ref{fig:noncentral_01_ferm}, in which the site $n=3$ is probed. We note that the mean occupation of the probed site is restricted to values below or equal to unity in the left panel, which is in agreement with the Pauli principle and the emerging unit population of the fermionic subspace. The right panel shows a simulation of the same system, where the measurements drive the system to the bosonic subspace, permitting the mean population of the probed site to reach values larger than unity.

The bottom panels of \ref{fig:noncentral_01_ferm} show that fermionic states reach a steady entanglement entropy of unity in our simulation, whereas the bosonic states present a persistently fluctuating noisy entanglement entropy. 
This difference can be attributed to the different requirements needed to enforce fermionic and bosonic symmetry.
A fermionic state for two distinguishable particles needs to have a minimal entanglement entropy of 1 ebit, corresponding to the antisymmetric state of the form $\frac{1}{\sqrt{2}}(\ket{\phi_1}\ket{\phi_2} - \ket{\phi_2}\ket{\phi_1})$. 
A higher entanglement entropy, $\mathcal{E}=2$, is possible for states of the form
\begin{equation}\label{eq:4s}
    \ket{\psi} = \frac{1}{2} \left(\ket{\phi_1}\ket{\phi_2} - \ket{\phi_2}\ket{\phi_1} + \ket{\phi_3}\ket{\phi_4} - \ket{\phi_4}\ket{\phi_3} \right).
\end{equation}
We do not observe states of this form in the long time limit because our measurement signal is able to resolve the different pairs of orthogonal states of such superpositions, and hence select a minimal fermionic state of the form $\frac{1}{\sqrt{2}}(\ket{\phi_1}\ket{\phi_2} - \ket{\phi_2}\ket{\phi_1})$. We note that probing of the central site does not resolve or affect odd states, and hence an initial state of the form \eqref{eq:4s}, where all $\ket{\phi_i}$ are odd states, will retain its entanglement entropy in the presence of probing. 

In the bosonic subspace, instead, double occupation allows the measurement to simultaneously probe the two atoms and affect their entanglement entropy.
These states can thus take the form of doubly occupied single particle states with $\mathcal{E}=0$, symmetrized products of orthogonal states with $\mathcal{E}=1$ and linear combinations of multiple product states with higher entanglement entropy. Measurements cause a persistent evolution among such states within the bosonic subspace, with a resulting fluctuating entanglement entropy.

\section{Three particles}
For three or more particles, the mathematical structure of quantum mechanics permits symmetries beyond bosons and fermions \cite{hartle1969,tichy-molmer2017}. The question of whether these symmetries can be generated and observed in a subsector of the Hilbert space is the central question of this Section.

\subsection{Representation theory and generalized symmetries}
We now review how symmetries beyond bosons and fermions can be understood in terms of \emph{irreducible representations} of the symmetric group $S_N$ \cite{hartle1969, tichy-molmer2017, sagan2000,zee2016}. 

Given a group $G$, a representation $r$ is a homomorphism
\begin{equation}
    r: G \longrightarrow GL(V)
    \label{eq:representation_def}
\end{equation}
where $GL(V)$ is the group of all invertible linear transformations on a vector space $V$. In particular, if $V=\mathbb{C}^d$, then $GL(V)$ is the group of all $d\times d$ invertible matrices with complex entries.
If we call $D^{(r)}(g)$ the matrix representing $g\in G$ in the representation $r$, we have
\begin{equation}
    D^{(r)}(g_1)D^{(r)}(g_2) = D^{(r)}(g_1g_2), \hspace{0.8cm} \forall g_1,g_2\in G
    \label{eq:representation_criterion}
\end{equation}
i.e. the representation preserves the multiplication structure of the group.

We define a representation $r: G \longrightarrow GL(V)$ as \emph{reducible} if there is a non-trivial subspace $U\subset V$ ($U\neq \{0\}, V$) such that
\begin{equation}
    D^{(r)}(g) U\subset U, \;\;\; \forall g\in G
\end{equation}
If $V$ contains no such subspace, then $r$ is called an \emph{irreducible representation}, or \emph{irrep}.

For finite groups, a reducible representation can be written as a direct sum of irreps $r = \oplus_i r_i$, with $r_i: G\longrightarrow GL(U_i)$. The vector space $V$ is similarly decomposed in \emph{irreducible subspaces} $V = \oplus_i U_i$. 

Different irreducible subspaces correspond to different symmetry properties under group action. In particular, irreducible subspaces under the action of $S_N$ encode different types of permutation symmetry of a tensor product.

We make this more concrete by considering a system of 3 particles. Given 3 distinct single-particle states $\ket{a}$, $\ket{b}$, $\ket{c}$, the set of possible permutations of the product state $\ket{\Psi}=\ket{a}\otimes\ket{b}\otimes\ket{c}$ forms a $3!=6$-dimensional vector space, that we call $W_\Psi$. Within this vector space we identify the bosonic subspace, spanned by the vector
\begin{equation}
\ket{\Psi_\.{bos}} = \ket{abc} + \ket{bac}+\ket{acb}+\ket{cba}+\ket{bca}+\ket{cab}
\label{eq:S3_psibos}
\end{equation}
and the fermionic subspace, spanned by 
\begin{equation}
\ket{\Psi_\.{ferm}} = \ket{abc} - \ket{bac}-\ket{acb}-\ket{cba}+\ket{bca}+\ket{cab}
\label{eq:S3_psiferm}
\end{equation}
The bosonic and fermionic subspaces are one-dimensional permutation-invariant subspaces: the bosonic state is unaffected by any permutation, and the fermionic one is unaffected up to a global sign change. These are the two most simple actions that the permutation group $S_N$ can perform on a tensor product space, and they are related to the trivial and sign representation of $\mathcal{S}_N$. They are the only two one-dimensional irreps of $S_N$.

We can see that \eqref{eq:S3_psibos} and \eqref{eq:S3_psiferm} do not span the entire $W_\Psi$.
Indeed, there are more permutation-invariant subspaces, related to higher-dimensional representations of $S_N$. Particles following these generalized symmetries have been recently referred to as  \emph{immanons}, because their state expansions have coefficients that are so-called immanants, which are a generalization of the Slater determinant and the permanent used for fermions and bosons \cite{tichy-molmer2017}.

As a property of representation theory \cite{zee2016}, an irrep $\lambda$ of dimension $d_\lambda$ is associated to $d_\lambda$ independent subspaces of $W_\Psi$, so that the sector associated to $\lambda$ has dimension $d_\lambda^2$.

For $\mathcal{S}_3$, in particular, there are three irreps: trivial (bosonic), sign (fermionic), and the \emph{standard representation}. The bosonic and fermionic sectors are one-dimensional, while the standard representation has a dimension $d_\.{st}=2$. There are therefore $2^2=4$ linearly independent vectors forming this immanonic subspace, and they form two distinct subspaces, which we may call $\mathbf{2_{st}}$, generated by \cite{hartle1969}
\begin{equation}
\begin{cases}
    &\ket{\psi_{st,1}} = \frac{1}{2} (\ket{abc}+\ket{bac}-\ket{cba}-\ket{cab})\\
    & \begin{aligned}
        \ket{\psi_{st,2}} = \frac{1}{2\sqrt{3}} (&\ket{abc}+\ket{bac}-2\ket{acb}+\\
        &+\ket{cba}-2\ket{bca}+\ket{cab})
    \end{aligned}
\end{cases}
    \label{eq:S3_psi_st}
\end{equation}
and $\mathbf{2_{st}'}$, generated by
\begin{equation}
\begin{cases}
    &\ket{\psi_{st,1}'} = \frac{1}{2} (\ket{abc}-\ket{bac}+\ket{cba}-\ket{cab})\\
    & \begin{aligned}
        \ket{\psi_{st,2}'} = \frac{1}{2\sqrt{3}} (&\ket{abc}-\ket{bac}+2\ket{acb}\\
        &-\ket{cba}-2\ket{bca}+\ket{cab})
    \end{aligned}
\end{cases}
\label{eq:S3_psi_st'}
\end{equation}
It can be seen that any permutation of the particles in a state belonging to $\mathbf{2_{st}}$ or $\mathbf{2_{st}'}$ yields a vector that still belongs to the same subspace.

An interesting consequence of $\mathbf{2_{st}}$ and $\mathbf{2_{st}'}$ having neither bosonic nor fermionic symmetry is that the corresponding immanons follow a \emph{partial Pauli principle} where single particle  states have a maximum allowed occupation (unlike bosons) but this occupation is higher than 1 (unlike fermions). For instance, if we consider non-distinct states, such as  $\ket{a}=\ket{b}$, and we substitute it into $\ket{\psi_{st,1}}$, we find the non-vanishing state (neglecting normalization) 
\begin{equation*}
    \ket{\psi_{st,1}} = 2\ket{aac}-2\ket{caa}\neq 0 \hspace{0.7cm} %\.{\Large \ding{51}}
\end{equation*}
whereas if $\ket{a}=\ket{b}=\ket{c}$:
\begin{equation*}
    \ket{\psi_{st,1}} = 2\ket{aaa}-2\ket{aaa} = 0 \hspace{0.7cm} %\.{\Large \ding{55}}
\end{equation*}
Therefore, $\ket{\psi}_{st,1}$ allows for double, but not for triple occupation of the same single particle state. 

Irreducible representations of the symmetric group are conveniently represented by Young diagrams, which are arrangements of boxes in which rows are symmetrized and columns are anti-symmetrized. They are associated to a broader formalism \cite{sagan2000}, which provides algorithms to easily calculate irrep properties such as the dimension and the character tables.

The Young diagrams for the irreps of $\mathcal{S}_3$ are:
\begin{equation*}
(1,1,1)=\vcenter{\hbox{{\scriptsize \ydiagram{1,1,1}}}}\,, 
\qquad (2,1) = \vcenter{\hbox{{\scriptsize\ydiagram{2,1}}}}\,, 
\qquad (3) = \vcenter{\hbox{{\scriptsize\ydiagram{3}}}} 
\end{equation*}
The $(1,1,1)$ diagram is thus completely anti-symmetric (fermionic) and $(3)$ is completely symmetric (bosonic), while $(2,1)$ is partially symmetric and partially anti-symmetric: its corresponding representation allows a maximum of 2 particles in one state and 1 particle in an orthogonal state. We call a particle following this latter symmetry a \emph{(2,1)-immanon}.

\subsection{Irrep projectors (symmetrizers)}
For the two-particle case, we have defined the bosonic and fermionic projectors \eqref{eq:P_bos_2part}-\eqref{eq:P_ferm_2part} in terms of the particle exchange operator. The definition was very easily given, because exchange is the only possible nontrivial permutation of two objects. For $N$ particles, we introduce the particle permutation operator $\hat{P}_\sigma$:
\begin{equation}
    \hat{P}_\sigma \, \ket{\psi_1}\otimes ... \otimes \ket{ \psi_n} = \ket{\psi_{\sigma_1}}\otimes...\otimes\ket{\psi_{\sigma_n}}
\end{equation}
where $\sigma\in \mathcal{S}_N$ is any possible permutation of $N$ objects. 

It can be shown \cite{tichy-molmer2017} that the projectors on the irreducible subspaces, or symmetrizers, can then be found as
\begin{equation}
    \hat{\mathcal{P}}^{(\lambda)} = \frac{d_\lambda}{N!}\sum_{\sigma\in S_N} \chi^{(\lambda)}(\sigma) \hat{P}_\sigma
    \label{eq:symmetrizer}
\end{equation}
where $\lambda$ is the irreducible representation and $d_\lambda$ is its dimension. $\chi^{(\lambda)}(\sigma)$ is the \emph{character function}, that encapsulates many properties of representations, and it is defined as the trace of the matrix representing the group element $\sigma$ in the representation $\lambda$. For $S_3$, in particular, the permutations fall into three possible conjugacy classes: $\sigma=I$ (identity), $\sigma=(12),(13),(23)$ (transpositions) and $\sigma=(123),(132)$ (3-cycles). The character is identical within each conjugacy class.

The tables below contain the character function for $\mathcal{S}_2$ and $\mathcal{S}_3$.
\begin{table}[H]
    \centering
    \begin{tabular}{c|c | c }
    $N=2$ & $\sigma= I$ & $\sigma = (12)$ \\ \hline
     $\chi^{\.{(2)}}$&  1 & 1 \\ \hline
     $\chi^{\.{(1,1)}}$ &   1 & -1
    \end{tabular}
    \caption{Character table of $S_2$}
    \label{tab:charactertable_s2}
\end{table}
\begin{table}[H]
    \centering
    \begin{tabular}{c|c | c | c}
    $N=3$ & $\sigma= I$ & $\sigma = (12)$ & $\sigma=(123)$ \\ \hline
     $\chi^{\.{(3)}}$&  1 & 1 & 1 \\ \hline
     $\chi^{\.{(1,1,1)}}$ &   1 & -1 & 1\\ \hline
     $\chi^{\.{(2,1)}}$& 2 & 0 & -1
    \end{tabular}
    \caption{Character table of $S_3$}
    \label{tab:charactertable_s3}
\end{table}

\begin{figure*}[!t]
    \centering
    \begin{subfigure}{0.32\linewidth}
        \includegraphics[width=\linewidth]{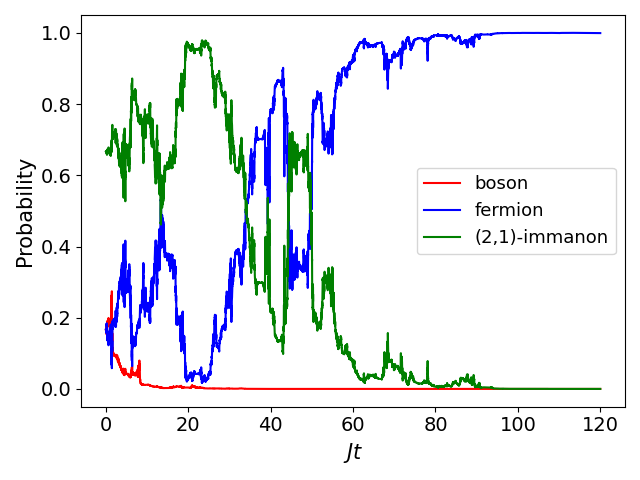}
    \end{subfigure}
    \begin{subfigure}{0.32\linewidth}
        \includegraphics[width=\linewidth]{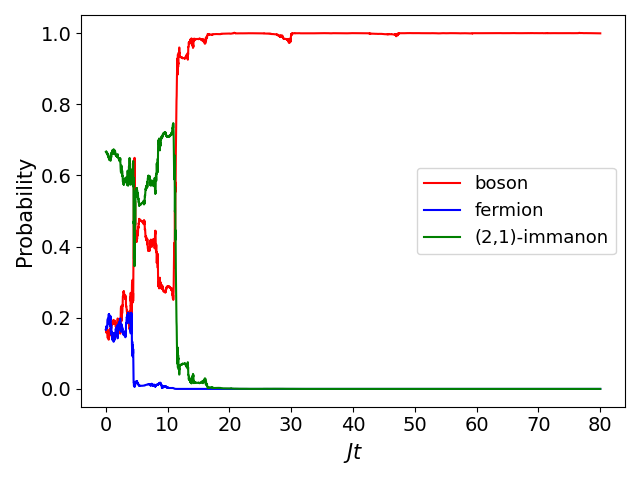}
    \end{subfigure}
    \begin{subfigure}{0.32\linewidth}
        \includegraphics[width=\linewidth]{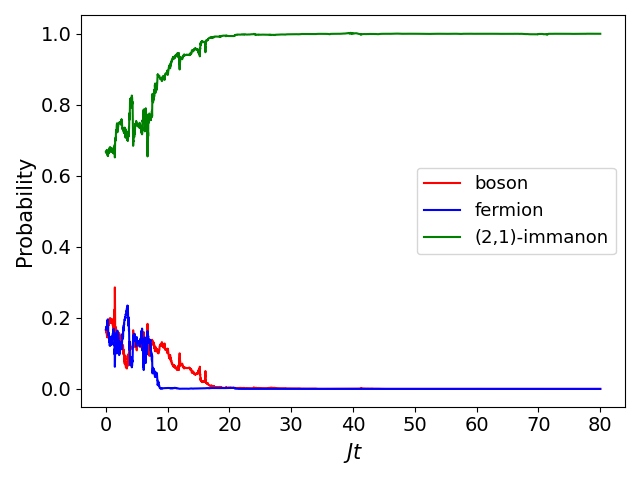}
    \end{subfigure}
    \begin{subfigure}{0.32\linewidth}
        \includegraphics[width=\linewidth]{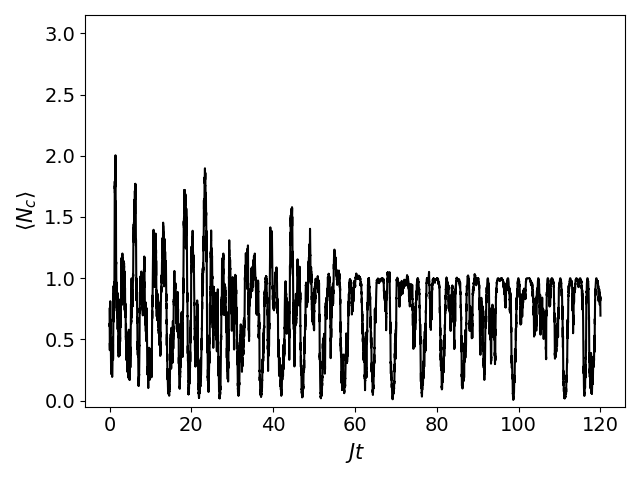}
    \end{subfigure}
    \begin{subfigure}{0.32\linewidth}
            \includegraphics[width=\linewidth]{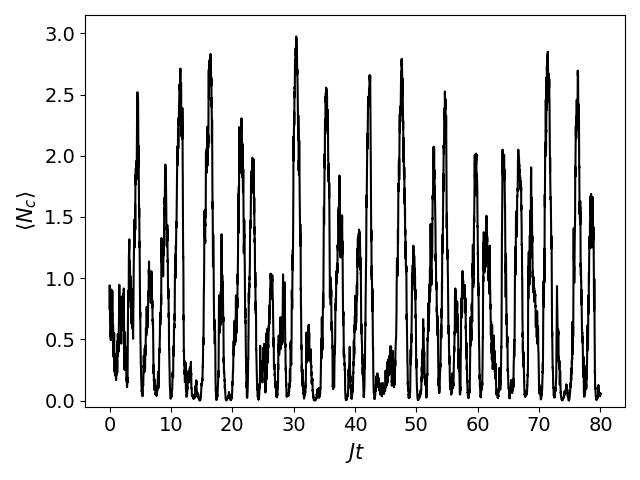}
    \end{subfigure}
    \begin{subfigure}{0.32\linewidth}
        \includegraphics[width=\linewidth]{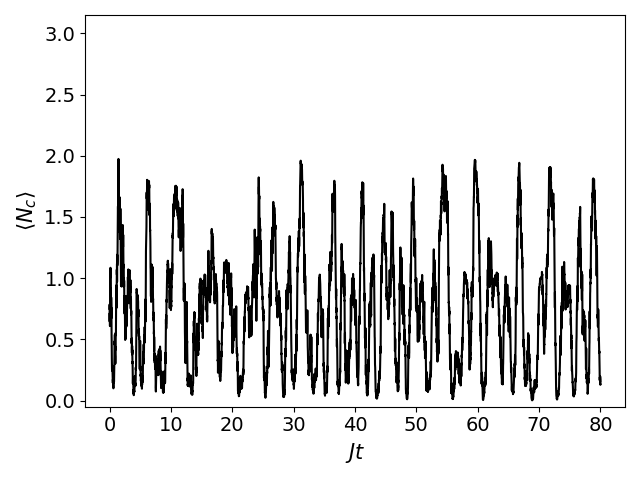}
    \end{subfigure}
    \begin{subfigure}{0.32\linewidth}
        \includegraphics[width=\linewidth]{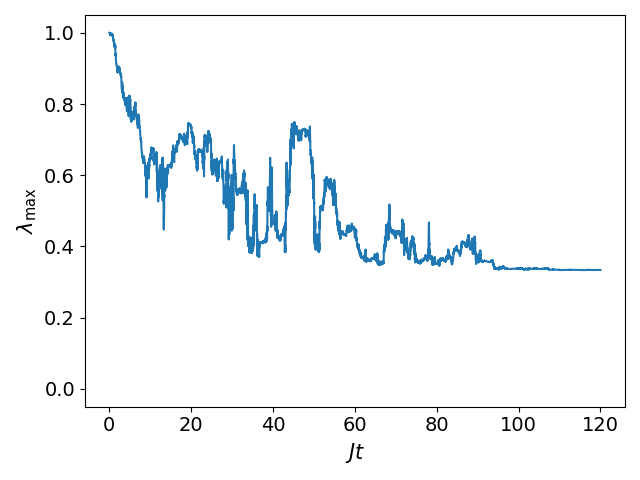}
    \end{subfigure}
    \begin{subfigure}{0.32\linewidth}
            \includegraphics[width=\linewidth]{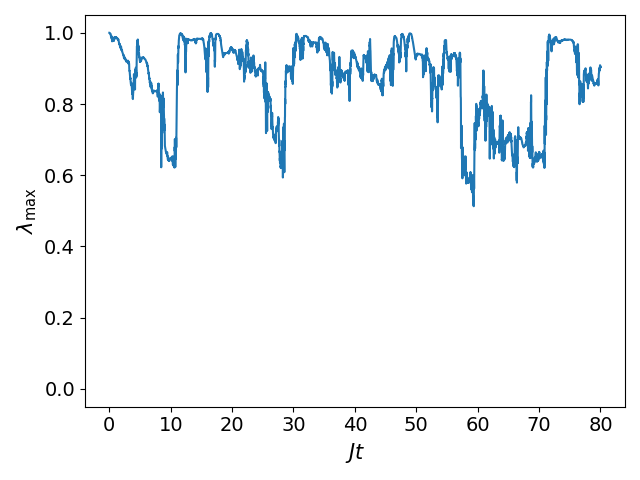}
    \end{subfigure}
    \begin{subfigure}{0.32\linewidth}
        \includegraphics[width=\linewidth]{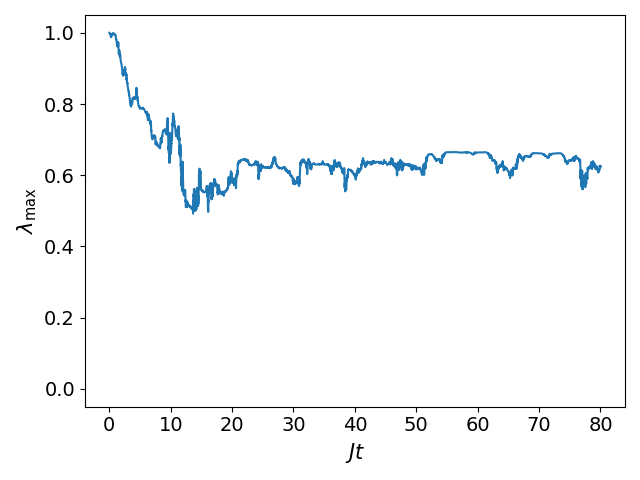}
    \end{subfigure}
    
    \caption{Fermionic, immanonic and bosonic states, obtained from $\ket{\psi(0)}=\ket{E_1}\ket{E_3}\ket{E_5}$, with central probing. The upper panels show the expectation values of the symmetry projectors, the central panels the occupation of the middle site and the lowest panels the maximum eigenvalue of the reduced density matrix for the first particle.}
\label{fig:fermionic_bosonic_immanonic}
\end{figure*}

\subsection{Selection of immanonic subspace by continuous probing}

The mathematical possibility of irreducible subspaces beyond bosons and fermions raises the question of whether these can be selected by probing, similarly to how the fermionic and bosonic subspace are selected for two-particle initially-separable states.

We consider the product state of three energy eigenstates with even symmetry, so that all particles are allowed to individually occupy the central site. We choose
\begin{equation}
    \ket{\psi(0)} = \ket{E_1}\ket{E_3}\ket{E_5} \label{eq:init_state_3part}
\end{equation}
with $L=7$ and probing on the central site ($p=4$).
The permutation symmetry of the state during the time evolution can be found by evaluating the expectation value of the symmetrizers. The initial state \eqref{eq:init_state_3part} is such that
\begin{equation}
    \braket{\hat{\mathcal{P}}^{\.{ bos}}} = \frac{1}{6}, \quad 
    \braket{\hat{\mathcal{P}}^{\.{ ferm}}} = \frac{1}{6}, \quad
    \braket{\hat{\mathcal{P}}^{\.{(2,1)}}} = \frac{2}{3}
\end{equation}
We therefore expect that the state be projected onto the bosonic subspace 1/6 of the times, the fermionic subspace 1/6 of the times, and the immanonic subspace 2/3 of the times. 
These statistics are confirmed by an ensemble of numerical trajectories we have observed, of which fig. \ref{fig:fermionic_bosonic_immanonic} illustrates an example for each selection.

The fermionic subspace is associated to an occupation number $\braket{N_p}$ fluctuating between 0 and 1, consistently with the Pauli exclusion principle, whereas for the bosonic subspace, the occupation number is allowed to take any value between 0 and 3. The most common outcome, occurring with a frequency of 2/3, is a value of $\braket{N_p}$ that fluctuates between 0 and 2, in accordance with the partial Pauli principle of the (2,1)-immanon statistics.

The observed dynamics of $\braket{N}_p$ only demonstrates that the (partial) Pauli principle applies to the probed, central site occupation. In order to verify it in every basis, we introduce $\lambda_\.{max}(t)$, defined as the largest eigenvalue of the reduced single particle density matrix $\rho_1(t)$. In our simulations, the largest eigenvalue of $\rho_1(t)$, $\rho_2(t)$ and $\rho_3(t)$ is found to be the same, providing the interpretation of $\lambda_\.{max}$ as the probability of the maximally-occupied single-particle mode, and $3\lambda_\.{max}(t)$ as the largest possible population in a certain mode at a given time.
Therefore, we expect the following inequalities must hold:
\begin{equation}
    3\lambda_\.{max} \leq 
    \begin{cases}
        &1, \hspace{0.5cm} \.{for fermions}\\
        &2, \hspace{0.5cm} \.{for (2,1)-immanons}\\
        &3, \hspace{0.5cm} \.{for bosons}
    \end{cases}
\end{equation}
This is, indeed, confirmed by our simulations, as illustrated in the lower panels of figure \ref{fig:fermionic_bosonic_immanonic}. We note that the largest eigenvalue for the reduced density matrix of the first particle (the pure state $\ket{E_1}$) starts at unity, but then gradually settles to values $\lambda_\.{max}= \frac{1}{3}$ for fermions and $\lambda_\.{max}\leq \frac{2}{3}$ for immanons. 

%Moreover, our simulations reveal that the subspaces thus found are also robust under dynamics and they are unaffected by any permutation-invariant Hamiltonian or measurement operators. In particular, changing the probed site or switching measurement off altogether does not affect the subspace.

\section{Conclusions}
The starting premise of this work was to show how measurements on dynamical quantum systems can reveal not only the fluctuations of the quantity measured but also other system observables and properties. 

A previously studied model with a particle moving on a lattice with a single weakly probed site \cite{blattmann2016} was extended to two and three particles, where probing heralds entanglement between the particles. Assuming different, but practically indistinguishable particles, we saw how probing permits selection of the permutation symmetry of the state, including the known bosonic and fermionic symmetries but also a third class of immanonic symmetry. 

Several works have recently explored the emergence of this type of parastatistics, in multiphoton interference \cite{kumar2025} and in quasiparticle excitations in spin models \cite{wang-hazzard2025}.
These and our results are not at variance with the fact that all elementary particles have so far been either fermions or bosons. Convincing arguments have been presented recently against any other possibilities, based on either quantum information arguments and the robustness of symmetry under composition of larger systems \cite{mekonnen2025} or thermodynamic constraints on the counting of microstates  \cite{zhou2025}.

In our study, the symmetry classification only applies to a subsector of the Hilbert space, accessed by the measurements and it is not enforced beyond this subsector.
We are thus not proposing the existence of new classes of real elementary or composite particles with non-standard symmetry. Instead, we point to the possibility of observing and simulating dynamics that obey such an emergent symmetry in  systems with a finite number of particles. 
Experimental implementations of such dynamics may be of foundational interest, and
the non-local correlations exhibited by the immanon subspaces  may make them of interest for quantum information protocols. 

\vspace{0.3cm}
\section{Acknowledgment}
This work was supported by the Danish National Research Foundation  (Center of Excellence “Hy-Q,” grant
number DNRF 139).

\newpage
\bibliography{Bib_paper}

@book{qip_bergou,
    author = {János A. Bergou and Mark Hillery and Mark Saffman},
    title = {Quantum information processing. Theory and implementations},
    publisher = {Springer},
    year = {2021},
    edition = {2nd}}

@article{qutip5,
  title = {QuTiP 5: The Quantum Toolbox in {Python}},
  author = {
    Lambert, Neill and Giguère, Eric and Menczel, Paul and Li, Boxi and
    Hopf, Patrick and Suárez, Gerardo and Gali, Marc and Lishman, Jake and
    Gadhvi, Rushiraj and Agarwal, Rochisha and Galicia, Asier and Shammah, Nathan and
    Nation, Paul and Johansson, J. R. and Ahmed, Shahnawaz and Cross, Simon and
    Pitchford, Alexander and Nori, Franco
  },
  journal = {Physics Reports},
  volume = {1153},
  pages = {1-62},
  year = {2026},
  issn = {0370-1573},
  doi = {10.1016/j.physrep.2025.10.001},
  url = {https://www.sciencedirect.com/science/article/pii/S0370157325002704},
}

@book{doob1953,
    author = {J. L. Doob},
    title = {Stochastic Processes},
    publisher = {Wiley},
    year = {1953}
}

@article{cao2019,
    author = {Xiangyu Cao and Antoine Tilloy and Andrea {De Luca}},
    title = {Entanglement in a fermion chain under continuous monitoring},
    journal = {SciPost Physics},
    volume = {7},
    number = {024},
    year = {2019},
    url = {https://scipost.org/SciPostPhys.7.2.024}
}

@article{kumar2025,
    author = {Shreya Kumar and Alex E Jones and Daniel Bhatti and Stefanie Barz},
    title = {Exchange Symmetry in Multiphoton Quantum Interference},
    journal = {Physical Review A},
    volume = {114},
    pages = {043702},
    year = {2026},
    url = {https://journals.aps.org/pra/abstract/10.1103/4thx-c9s7}
}

@article{zhou2025,
    author = {Chi-Chun Zhou and Shuai A. Chen and Yu-Zhu Chen and Yao Shen and Fu-Lin Zhang and Wu-Sheng Dai},
    title = {Quantum Statistics Forbids Particle Exchange
Statistics beyond Bosons and Fermions in 3D},
    journal = {arXiv:2505.17361v3 [quant-ph]},
    year = {2025},
    url = {https://arxiv.org/abs/2505.17361}
}

@article{Gorshkov2010,
  title = {Two-orbital SU(N) magnetism with ultracold alkaline-earth atoms},
  author = {A. V. Gorshkov and M. Hermele and  V. Gurarie and C. Xu and P. S. Julienne  and J. Ye and P. Zoller and  E. Demler and M. D. Lukin and A. M. Rey},
  journal = {Nature Physics},
  volume = {6},
  pages = {289--295},
  year = {2010},
  url = {https://www.nature.com/articles/nphys1535}
}

@article{hartle1969,
    author = {James B. Hartle and John R. Taylor},
    title = {Quantum Mechanics of Paraparticles},
    journal = {Physical Review},
    volume = {178},
    number = {5},
    pages = {2043-2051},
    year = {1969},
    url = {https://journals.aps.org/pr/abstract/10.1103/PhysRev.178.2043}
}

@article{tichy-molmer2017,
    author = {Malte C. Tichy and Klaus Mølmer},
    title = {Extending bosons and fermions beyond pairwise exchange symmetry},
    journal = {Physical Review A},
    volume = {96},
    pages = {022119},
    year = {2017},
    url = {https://journals.aps.org/pra/abstract/10.1103/PhysRevA.96.022119}
}

@article{mekonnen2025,
    author = {Manuel Mekonnen and Thomas D. Galley and Markus P. Müller},
    title = {Invariance under quantum permutations rules out parastatistics},
    journal = {Nature Communications},
    volume = {17},
    pages = {6947},
    year = {2026},
    url = {https://www.nature.com/articles/s41467-026-73064-6}
}

@book{sagan2000,
    author = {Bruce E. Sagan},
    title = {The Symmetric Group. Representations, Combinatorial Algorithms and Symmetric Functions},
    publisher = {Springer},
    year = {2000},
    edition = {2nd edition}
}

@book{zee2016,
    author = {Anthony Zee},
    title = {Group theory in a nutshell for physicists},
    publisher = {Princeton University Press},
    year = {2016}
}

@article{cabrillo1999,
    author = {C. Cabrillo and J. I. Cirac and P. García-Fernández and P. Zoller},
    title = {Creation of entangled states of distant atoms by interference},
    journal = {Physical Review A},
    volume = {59},
    number = {2},
    pages = {1025},
    year = {1999},
    url = {https://journals.aps.org/pra/abstract/10.1103/PhysRevA.59.1025}
}

@article{sherson2010,
    author = {Jacob F. Sherson and Christof Weitenberg and Manuel Endres and Marc Cheneau and Immanuel Bloch and Stefan Kuhr},
    title = {Single-atom-resolved fluorescence imaging of an
            atomic Mott insulator},
    journal = {Nature},
volume	= {467},
issue	= {7311},
pages	= {68--72},
    year = {2010},
    url = {https://www.nature.com/articles/nature09378}
}

@article{bakr2009,
    author = {Waseem S. Bakr and Jonathon I. Gillen and Amy Peng and Simon Fölling and Markus Greiner},
    title = {A quantum gas microscope for detecting single atoms in a Hubbard-regime optical lattice},
    journal = {Nature},
    volume	= {462},
    issue	= {7269},
    pages	= {74--77},
    year = {2009},
    url = {https://www.nature.com/articles/nature08482}
}

@article{hammerer2010,
    author = {Klemens Hammerer and Anders S. Sørensen and Eugene S. Polzik},
    title = {Quantum interface between light and atomic ensembles},
    journal = {Reviews of Modern Physics},
    volume = {82},
    pages = {1041--1093},
    year = {2010},
    url = {https://journals.aps.org/rmp/abstract/10.1103/RevModPhys.82.1041}
}

@article{yamamoto2017,
    author = {Ryuta Yamamoto and Jun Kobayashi and Kohei Kato and Takuma Kuno and Yuto Sakura and Yoshiro Takahashi},
    title = {Site-resolved imaging of single atoms with a Faraday quantum gas microscope},
    journal = {Physical Review A},
    volume = {96},
    pages = {033610},
    year = {2017},
    url = {https://journals.aps.org/pra/abstract/10.1103/PhysRevA.96.033610}
}

@article{blattmann2016,
    author = {Ralf Blattmann and Klaus Mølmer},
    title = {Conditioned quantum motion of an atom in a continuously monitored one-dimensional lattice},
    journal = {Physical Review A},
    volume = {93},
    pages = {052113},
    year = {2016},
    url = {https://journals.aps.org/pra/abstract/10.1103/PhysRevA.93.052113}
}

@phdthesis{valiente2010,
    author = {Manuel Valiente Cifuentes},
    title = {Few Quantum Particles on One Dimensional Lattices},
    school = {Humboldt University of Berlin},
    year = {2010}
}

@article{jakschzoller2005,
title	= {The cold atom Hubbard toolbox},
author	= {D. Jaksch and P. Zoller},
journal	= {Annals of Physics},
volume = {315},
issue = {1},
year	= {2005},
url = {https://www.sciencedirect.com/science/article/abs/pii/S0003491604001782}
}

@book{Jacobs2014,
    author = {Kurt Jacobs},
    title = {Quantum Measurement Theory and its Applications},
    publisher = {Cambridge University Press},
    year = {2014}
}

@article{anton-molmer,
author = {A. L. Andersen and Klaus Mølmer},
title = {Quantum non-demolition measurements of moving target states},
year = {2022},
journal = {Physical Review Letters},
volume= {129},
pages = {120402},
url = {https://journals.aps.org/prl/abstract/10.1103/PhysRevLett.129.120402}
}

@article{wang-hazzard2025,
    author = {Zhiyuan Wang and Kaden R. A. Hazzard},
    title = {Particle exchange statistics beyond fermions and bosons},
    journal = {Nature},
    volume = {637},
    pages = {314--318},
    year = {2025},
    url = {https://www.nature.com/articles/s41586-024-08262-7}
}

\end{document}